\documentclass[runningheads]{llncs}

\usepackage{hyperref}
\usepackage{color, colortbl, soul}
\definecolor{lightyellow}{rgb}{1,1,1}
\definecolor{lightGray}{gray}{0.9}
\sethlcolor{lightyellow}
\definecolor{yellow}{rgb}{1,1,1}
\usepackage[cmex10]{amsmath}
\usepackage{balance}
\usepackage{makeidx}
\usepackage{amssymb}
\usepackage{latexsym}
\usepackage{graphicx}
\usepackage{comment}
\usepackage{booktabs}
\usepackage{multirow}
\usepackage{framed}
\usepackage{longtable}
\usepackage{subcaption}
\usepackage{makecell}
\usepackage{soul}
\definecolor{Gray}{gray}{0.75}
\definecolor{LGray}{gray}{0.95}

\begin{document}
\title{Do you really follow them? \\Automatic detection of credulous Twitter users\thanks{Partially supported by the European Union's Horizon 2020 programme (grant agreement No. 830892, SPARTA) and by IMT Scuola Alti Studi Lucca: Integrated Activity Project TOFFEe `TOols for Fighting FakEs'.
It has also benefited from the computing resources (ULITE) provided by the IT division of LNGS in L'Aquila.}}

\author{Alessandro Balestrucci \inst{1}
\and
Rocco~De~Nicola\inst{2} \and
\\Marinella~Petrocchi\inst{2,3} \and
Catia~Trubiani\inst{1}
}

\authorrunning{A. Balestrucci, R. De~Nicola, M. Petrocchi, C. Trubiani}
\institute{
Gran Sasso Science Institute, L'Aquila, Italy \\
\email{\{alessandro.balestrucci, catia.trubiani\}@gssi.it}\and
IMT School for Advanced Studies, Lucca, Italy,
\email{rocco.denicola@imtlucca.it}\\ \and
Institute of Informatics and Telematics (IIT-CNR), Pisa, Italy\\
\email{marinella.petrocchi@iit.cnr.it}
}
\maketitle
\begin{abstract}

Online Social Media represent a pervasive source of information able to reach a huge audience.
Sadly, recent studies show how online social bots (automated, often malicious accounts, populating social networks and mimicking genuine users) are able to amplify the dissemination of (fake) information by orders of magnitude.
Using Twitter as a benchmark, in this work we focus on what we define \textit{credulous} users, i.e., human-operated accounts with a high percentage of bots among their \textit{followings}. Being more exposed to the harmful activities of social bots, credulous users may run the risk of being more influenced than other users; even worse, although unknowingly, they could become spreaders of misleading information (e.g., by retweeting bots). We design and develop a supervised classifier to automatically recognize \textit{credulous} users. The best tested configuration achieves an accuracy of 93.27\% and AUC-ROC of 0.93, thus leading to positive and encouraging results. 

\end{abstract}
\section{Introduction}

The diffusion of information on Social Media is often supported by automated accounts, controlled totally or in part by computer algorithms, called bots. Unfortunately, a dominant and worrisome use of automated accounts is far from being benign: malicious bots are purposely created to distribute spam, sponsor public characters and, ultimately, induce a bias within the public opinion~\cite{ferrara2016rise}. Especially, their malicious activities are of high efficacy when performed on a targeted audience~\cite{bovet2019influence,bastos2019brexit} to, e.g., generate misconception or encourage hate campaigns~\cite{ChatzakouKBCSV17}. Recent work in~\cite{shao2018spread,yang2019arming} demonstrate that bots are particularly active in spreading low credibility content and amplifying their significance. Moreover, human-operated accounts contribute to the diffusion of disinformation by, e.g., retweeting and/or liking fake content. 

In a previous work~\cite{balestrucci2019credulous}, the authors shed light on so called \textit{credulous} Twitter users assuming, with a harmless abuse of language, that they refer to human-operated accounts with a high percentage of bots as friends.
 Unlike~\cite{balestrucci2019credulous}, where the authors performed an analysis involving the friends of a set of human-operated accounts - a highly time consuming task - here we design and develop a classifier to find out credulous Twitter users, by considering a number of features that do not take the friendship with bots into account. 
 Starting by considering a set of features commonly employed in the literature to detect bots~\cite{Varol17,Cresci15fame}, we end up with a lightweight classifier, in terms of costs for gathering the data needed for the feature engineering phase. The classification performance achieves very encouraging results -- an accuracy of 93.27\% and an \textit{AUC} (Area Under the ROC curve) equal to 0.93.

We believe that automatically detecting credulous users is  a promising
line of research.
Such an investigation could help researchers to: 
1. better understand the characteristics of those users more polarized and/or more willing to be influenced;
2. unveil low-credibility and/or deceptive content and limite their online
diffusion;
3. devise alternative strategies for bot detection by concentrating
the analysis on the friends of credulous users;
4. improve the users' awareness about threats to data trustworthiness.

The following section presents the approach for the automatic detection of credulous Twitter users, while  Section~\ref{sec:ExpRes} presents the experimental results. Section~\ref{sec:discussion} discusses the outcome and suggests further investigations. Section~\ref{sec:relatedWork} presents related work in the area,  arguing on the differences, the contributions and the novelty w.r.t. our work. Section~\ref{sec:conclusions} concludes the paper. 

\section{The approach}
\label{sec:approach}
\subsection{Datasets}
\label{subsec:datasets}
We consider three publicly available datasets\footnote{Bot Repository Datasets: \url{https://goo.gl/87Kzcr}}: \texttt{CR15}~\cite{Cresci15fame},
\texttt{CR17}~\cite{Cresci17paradigm}, and
\texttt{VR17}~\cite{Varol17}.
From the merging of these three datasets, we obtain a unique labeled dataset (human-operated/bot) of 12,961 accounts - 7,165 bots and 5,796 humans. We use this dataset to train a bot detector, as described in Section~\ref{subsec:BotDetector}. To this end, we
use the Java Twitter API\footnote{Twitter API: \url{https://goo.gl/njcjr1}}, and for each account we collect: tweets (up to 3,200), mentions (up to 100), IDs of friends and followers (up to 5k). 

The identification of credulous users follows the approach presented in~\cite{balestrucci2019credulous}. To this end, we need to detect the amount of bots which are friends of the 5,796 human-operated accounts. Due to the rate limits of the Twitter APIs and to the huge amount of friends possibly belonging to these human-operated accounts, we consider only those accounts with a list of friends lower than or equal to 400~\cite{balestrucci2019credulous}.  
This leads to a dataset of 2,838 human-operated accounts, namely \texttt{Humans2Consider} hereafter. By crawling the data related to their friends, we overall acquire information related to 421,121 Twitter accounts.

\subsection{Bot Detection}
\label{subsec:BotDetector}

A bot detection phase is required to discriminate bots and genuine accounts in the dataset of selected friends. 
The literature offers a plethora of successful approaches~\cite{ferrara2016rise}; however, also due to the capabilities of evolved spambots to evade detection~\cite{Cresci17paradigm}, the performances of the diverse techniques degenerate over time~\cite{minnich2017botwalk}. 
Furthermore, some bot detectors are available online, but not fully usable due to restrictions in their terms of use, see, e.g.,~\cite{chavoshi2016debot}. To overcome these issues, we design and develop a supervised approach, which mixes features from popular scientific work and novel features here introduced. 

Regarding the features, we consider two sets.
The first one derives from Botometer~\cite{Varol17}, a popular bot detector\footnote{\url{https://botometer.iuni.iu.edu/}}. In addition to the original Botometer features~\cite{Varol17}, we also include: the CAP\footnote{Complete Automation Probability: \url{https://tinyurl.com/yxp3wqzh}} (Complete Automation Probability) score, 
the Scores\footnote{English/Universal Score: \url{https://tinyurl.com/y2skbmqc}}, the number of tweets and mentions; we call \textit{Botometer+} this augmented set of features.
The second feature set is inherited from~\cite{Cresci15fame}, where a classifier was designed to detect fake Twitter followers.
 We use almost all their \textit{ClassA} features\footnote{ClassA features require only information available in the profile of the account~\cite{Cresci15fame}.}, except the one about duplicated pictures, because it was not possible for us to verify whether 
the same profile picture was used twice; we call \textit{ClassA-} this reduced set of features.
The conjunction of the two sets of features is referred in the following as \textit{ALL\_features}.

We use 19 learning algorithms to train our classifier (with a 10-fold cross validation) and we compare their classification capabilities with respect to 
the three feature sets (\textit{Botometer+}, \textit{ClassA-} and \textit{ALL\_features}). 
The classification performances are evaluated according to:
percentage of \textit{accuracy}, \textit{precision}, \textit{recall}, F-measure (\textit{F1}), and Area Under the ROC Curve (\textit{AUC}).
On the most accurate classifier, Hyper-Parameter tuning is performed. The tuned classifier is then used to label the friends of the \texttt{Humans2Consider} dataset (see Section~\ref{subsec:datasets}). 

\subsection{Identification of Credulous Twitter Users}
\label{subsec:CredIden}

The identification of credulous users can be performed with multiple strategies, since there are various aspects that may contribute to spot those users more exposed to the malicious activities of bots. 
In our previous work~\cite{balestrucci2019credulous}, we introduced a set of rules to discern whether a genuine user is a credulous one. These rules allow to rank users by relying on the ratio of bots over the user's list of friends. Here, we inherit these rules to rank the users in our dataset (see Section \ref{subsec:datasets}), but further ranking strategies can be also considered. Our goal is to build a \emph{ground truth} of credulous users to derive an assessed characterization of these accounts. 
Applying the approach defined in~\cite{balestrucci2019credulous}, we identified as \textit{credulous} 316 users in \texttt{Humans2Consider}.
This constitutes the input data for the next step.
We note that the approach in~\cite{balestrucci2019credulous} is very expensive in terms of data gathering. For example, for the investigated dataset, it requires 421k users' account information and 833 million of tweets. 

\subsection{Classification of Credulous Twitter Users}
\label{subsec:CredClass}

Goal of this phase is to build a decision model to automatically classify a Twitter account as credulous or not. As ground-truth, we consider the 316 accounts identified as credulous according to the process described in Section~\ref{subsec:CredIden}.

We experiment the same learning algorithms and the same feature sets considered in Section~\ref{subsec:BotDetector}, with 10 cross-fold validation. 
However, for credulous users classification, the learning algorithms take as input a very unbalanced dataset: we have 2,838 human-operated accounts (see Section~\ref{subsec:datasets}) and, among them, 316 have been identified as credulous accounts (see Section~\ref{subsec:CredIden}).
To avoid working with unbalanced datasets, 
we split the sets of not credulous users into smaller portions, equal to the number of credulous users.
We randomly select a number of not credulous users equal to the number of credulous ones; then, we unify these instances in a new dataset (hereinafter referred to as \textit{fold}). Then, we repeat this process on previously un-selected sets, until there are no more not credulous instances. Such procedure has been inspired by the \textit{under-sampling iteration} methodology, for strongly unbalanced datasets~\cite{lee2015iterative}.
Each learning algorithm is trained on each fold. 
To evaluate the classification performances on the whole dataset, and not just on individual folds, we compute the average of the single performance values, for each evaluation metric. 

\section{Experimental Results}
\label{sec:ExpRes}

All the experiments are performed with Weka~\cite{witten2016data}, a tool providing the implementation of several machine learning algorithms. In the following, we present the main results obtained for bot detection and credulous classification, all the details are publicly available: \url{https://tinyurl.com/y4l632g5}. 

The first column of Tables~\ref{table:classifiers} and~\ref{table:CredClassifiers753} shows the set of features considered for learning (i.e., \textit{ALL\_features}, \textit{Botometer+}, \textit{ClassA-}, see Section \ref{subsec:BotDetector}). The second column reports a subset of the
adopted machine learning algorithms whose name is abbreviated according to the Weka's notation and reported in the following: 

         \begin{framed}
         \noindent IBk: K-nearest neighbours~\cite{Aha1991}, NB: Naive Bayes~\cite{john1995estimating}, 
         SMO: Sequential Minimal Optimization~\cite{Platt1998}, JRip: RIPPER~\cite{cohen1995fast},
         MLP: Multi-Layer Perceptron~\cite{pal1992multilayer}, RF: Random Forest~\cite{Breiman2001}, REP: Reduced-Error Pruning~\cite{quinlan1987simplifying}, 1R~\cite{Holte1993} 
         \end{framed}

The remaining columns report the evaluation metrics mentioned above. 
\\
    \begin{center}
    \begin{longtable}[htbp]
    {p{0.5cm} p{0.1cm} p{1.0cm} p{1.7cm} p{1.7cm} p{1.5cm} p{1.5cm} p{1.5cm}}\\
    & & & \multicolumn{5}{c}{\textit{evaluation metrics}}\\
    \cline{4-8}
    & & \textit{alg} & \multicolumn{1}{r}{\textit{accuracy}} & \multicolumn{1}{r}{\textit{precision}} & \multicolumn{1}{r}{\textit{recall}} & \multicolumn{1}{r}{\textit{F1}}  
    & \multicolumn{1}{r}{\textit{AUC}} \\
    \hline
     \multicolumn{1}{l}{\multirow{3}{*}{\textit{ALL\_features}}} 
    & & SMO & \multicolumn{1}{r}{98.04} & \multicolumn{1}{r}{0.98} & \multicolumn{1}{r}{0.98} & \multicolumn{1}{r}{0.98} & 
    \multicolumn{1}{r}{0.98} \\
    & & JRip & \multicolumn{1}{r}{97.92} & \multicolumn{1}{r}{0.99} & \multicolumn{1}{r}{0.98} & \multicolumn{1}{r}{0.98} & 
    \multicolumn{1}{r}{0.99} \\
    & & \textbf{RF} & \multicolumn{1}{r}{\textbf{98.33}} & \multicolumn{1}{r}{\textbf{0.99}} & \multicolumn{1}{r}{\textbf{0.98}} & \multicolumn{1}{r}{\textbf{0.98}} & 
    \multicolumn{1}{r}{\textbf{1.00}} \\
    \hline
     \multicolumn{1}{l}{\multirow{3}{*}{\textit{Botometer+}}}
    & & SMO & \multicolumn{1}{r}{97.64} & \multicolumn{1}{r}{0.98} & \multicolumn{1}{r}{0.98} & \multicolumn{1}{r}{0.98} & 
    \multicolumn{1}{r}{0.98} \\
    & & JRip & \multicolumn{1}{r}{97.61} & \multicolumn{1}{r}{0.98} & \multicolumn{1}{r}{0.97} & \multicolumn{1}{r}{0.98} & 
    \multicolumn{1}{r}{0.98} \\
    & & \textbf{RF} & \multicolumn{1}{r}{\textbf{97.97}} & \multicolumn{1}{r}{\textbf{0.98}} & \multicolumn{1}{r}{\textbf{0.98}} & \multicolumn{1}{r}{\textbf{0.98}} & 
    \multicolumn{1}{r}{\textbf{1.00}} \\
    \hline
    \multicolumn{1}{l}{\multirow{3}{*}{\textit{ClassA-}}} 
    & & IBk & \multicolumn{1}{r}{91.03} & \multicolumn{1}{r}{0.91} & \multicolumn{1}{r}{0.93} & \multicolumn{1}{r}{0.92} & 
    \multicolumn{1}{r}{0.91} \\
    & & JRip & \multicolumn{1}{r}{94.38} & \multicolumn{1}{r}{0.96} & \multicolumn{1}{r}{0.94} & \multicolumn{1}{r}{0.95} & 
    \multicolumn{1}{r}{0.96} \\
    & & \textbf{RF} & \multicolumn{1}{r}{\textbf{95.84}} & \multicolumn{1}{r}{\textbf{0.98}} & \multicolumn{1}{r}{\textbf{0.95}} & \multicolumn{1}{r}{\textbf{0.96}} & 
    \multicolumn{1}{r}{\textbf{0.99}} \\
    \hline
    & &  &  &  &  &  &  \\
    \caption{Results for bot detection}
    \label{table:classifiers}
    \end{longtable}
    \vspace{-0.5cm}
    \end{center}
%
\begin{center}
\begin{longtable}[htbp]
{p{0.5cm} p{0.1cm} p{1.0cm} p{1.7cm} p{1.7cm} p{1.5cm} p{1.5cm} p{1.5cm}}
& & & \multicolumn{5}{c}{\textit{evaluation metrics}}\\
\cline{4-8}
& & \textit{alg} & \multicolumn{1}{r}{\textit{accuracy}} & \multicolumn{1}{r}{\textit{precision}} & \multicolumn{1}{r}{\textit{recall}} & \multicolumn{1}{r}{\textit{F1}} 
& \multicolumn{1}{r}{\textit{AUC}} \\
\hline
\multicolumn{1}{l}{\multirow{3}{*}{\textit{ALL\_features}}} & & IBk & \multicolumn{1}{r}{89.69} & \multicolumn{1}{r}{0.74} & \multicolumn{1}{r}{0.73} & \multicolumn{1}{r}{0.90} &  \multicolumn{1}{r}{0.96} \\
& &  \textbf{1R}  & \multicolumn{1}{r}{\textbf{93.27}} & \multicolumn{1}{r}{\textbf{0.99}} & \multicolumn{1}{r}{\textbf{0.88}} & \multicolumn{1}{r}{\textbf{0.93}} &  \multicolumn{1}{r}{\textbf{0.93}} \\
& & REP & \multicolumn{1}{r}{93.07} & \multicolumn{1}{r}{0.99} & \multicolumn{1}{r}{0.88} & \multicolumn{1}{r}{0.93} &   \multicolumn{1}{r}{0.94} \\
\hline
 \multicolumn{1}{l}{\multirow{3}{*}{\textit{Botometer+}}}  & & IBk & \multicolumn{1}{r}{65.03} & \multicolumn{1}{r}{0.61} & \multicolumn{1}{r}{0.60} & \multicolumn{1}{r}{0.63} &   \multicolumn{1}{r}{0.70} \\
& & JRip & \multicolumn{1}{r}{66.42} & \multicolumn{1}{r}{0.67} & \multicolumn{1}{r}{0.67} & \multicolumn{1}{r}{0.66} &   \multicolumn{1}{r}{0.67} \\
& & \textbf{RF} & \multicolumn{1}{r}{\textbf{67.81}} & \multicolumn{1}{r}{\textbf{0.68}} & \multicolumn{1}{r}{\textbf{0.69}} & \multicolumn{1}{r}{\textbf{0.68}} &   \multicolumn{1}{r}{\textbf{0.73}} \\
\hline
\multicolumn{1}{l}{\multirow{3}{*}{\textit{ClassA-}}} & & IBk & \multicolumn{1}{r}{92.59} & \multicolumn{1}{r}{0.74} & \multicolumn{1}{r}{0.73} & \multicolumn{1}{r}{0.92} &   \multicolumn{1}{r}{0.97} \\
& &  \textbf{1R} & \multicolumn{1}{r}{\textbf{93.27}} & \multicolumn{1}{r}{\textbf{0.99}} & \multicolumn{1}{r}{\textbf{0.88}} & \multicolumn{1}{r}{\textbf{0.93}} &   \multicolumn{1}{r}{\textbf{0.93}} \\
& & REP & \multicolumn{1}{r}{93.09} & \multicolumn{1}{r}{0.98} & \multicolumn{1}{r}{0.88} & \multicolumn{1}{r}{0.93} &   \multicolumn{1}{r}{0.95} \\
\hline
& &  &  &  &  &  & \\
\caption{\vspace{-0.2cm}Results for credulous detection}
\label{table:CredClassifiers753}
\end{longtable}
\end{center}
\vspace{-0.48cm}

Regarding bot detection, Table~\ref{table:classifiers} shows that all the machine learning algorithms well behave, regardless of the feature set. 
Random Forest is the one that performs best. When the set {\it ALL\_features} is used, the results are:
accuracy = 98.33\%, F1 = 0.98 and AUC = 1.00; and after the tuning phase, we obtain a final accuracy = 98.41\%. 

Table~\ref{table:CredClassifiers753} shows that \textit{ALL\_features} and \textit{ClassA-} have good and quite similar classification performances, contrary to \textit{Botometer+}. Both \textit{ALL\_features} and \textit{ClassA-} demonstrate their efficacy to discriminate credulous users.  
On the contrary, the \textit{Botometer+}'s features properly work for bot detection tasks only.
Going into deeper details, in Table \ref{table:CredClassifiers753} we can notice that the 1R algorithm obtains the best accuracy percentage (93.27\% with $\sigma= 3.22$) and F-score (0.93), but not the highest AUC (0.93). 
It is worth noting that the values of the 1R algorithm are exactly the same when considering \textit{ALL\_features} and \textit{ClassA-}. This means that the algorithm selects \textit{ClassA-}'s features only, the ones from \textit{Botometer+} are useless in this case. This is a relevant result since we recall that \textit{ClassA-} features refer to the profile of accounts and it is less expensive to collect them.

\section{Discussion}
\label{sec:discussion}
\vspace{-0.5em}
The results in Table~\ref{table:CredClassifiers753} show the capability of our approach to automatically discriminate those Twitter users with a large number of bots as friends, namely credulous, {\it without explicitly considering the features of the latter, which would imply a very high cost in terms of data gathering}. 
To better understand this point, we recall that the approach in~\cite{balestrucci2019credulous} for the identification of credulous users needs to crawl a large amount of data, due to the necessity of extending the analysis to the friends of a Twitter account. In the specific case under investigation, this means to retrieve information for more than 400k user accounts, 11 millions of tweet mentions, and more than 820 millions of tweets. As opposite, the credulous detector here proposed requires to gather the profile information of 2,838 accounts only.
The classification performances are really promising, with the best accuracy  93.27\%, best F1 0.93, best AUC 0.93. We remark that such results have been achieved by relying on so called {\it ClassA-} features only, i.e., features extracted from the account profile. It is peculiar how the features useful to discriminate credulous genuine accounts are features belonging to the account profile only. This preliminary result calls for three further investigations: 1. to 
compare the range of values assumed by these features when detecting credulous accounts with the one assumed to detect social bots (as in~\cite{Cresci15fame}); 2. to explore the reason why more complex features (such as the ones of Botometer) do not seem to give good  results to find credulous users; 3. to perform a deeper analysis on the importance of each specific feature when discriminating credulous users, by means, e.g., of Principal Component Analysis~\cite{witten2016data}).

Finally, even if the design of a bot detector is not the primary goal of this paper, but only a mean through which we obtain the ground-truth for training the credulous user classifier, we notice that, compared to the performances reported in~\cite{Cresci17paradigm,yang2019arming}, our bot detector achieves very good classification performances.
This strengthens the robustness of the ground truth obtained in Section~\ref{subsec:CredIden}, since the friends' nature evaluation is assessed by means of a very accurate classifier.

\section{Related Work}
\label{sec:relatedWork}
\vspace{-0.5em}
Our work is related to all those approaches that investigate peculiar features of social networks users. 
We discuss the ones we find more relevant for our approach, with the caveat that the presented literature review is far from being exhaustive. 

A survey on  users' behaviour in social networks is proposed in~\cite{jin2013understanding}: it is remarked that the recipients of shared information should be chosen, in a more precautionary way, by taking into account more real-life relationships and less virtual links. Our approach works exactly in the direction of enhancing the awareness of users, by classifying the ones more exposed to attacks of social bots.

Information spreading on Twitter
is investigated in~\cite{DBLP:journals/corr/MonstedSFL17}, where the authors demonstrate that the probability of spreading a given piece of information is higher when promoted by multiple sources.
This supports our attempt to analyze the percentage of bots within the friends of human-operated Twitter accounts, as a symptom for being more tempted to disinformation.

In~\cite{amato2018recognizing}, human behaviour on Facebook is analyzed by building graphs that capture sequence of activities. Behavioural patterns that do not match any of the known benign models likely signal malicious objectives. Similarly, the realization of a classifier to automatically recognize credulous users is the first step to derive their sequence of activities and, hopefully, peculiar behavioral patterns. 

In~\cite{cresci2017humans,gilani2019large}, a behavioural analysis of bots and humans on Twitter is performed, to draw fundamental differences between the two groups. Specifically, the former demonstrates how, despite a higher level of synchronization characterizing bot accounts, the human behaviour on Twitter is far from being random. The latter defines a `credibility score' as a measure of how many tweets by bots are present in the timeline of an account. 
Our work supports the discrimination of credulous users and it may lead to a deeper characterization of human accounts.

To the best of our knowledge, few research explores ways to automatically recognize those Twitter users susceptible to attacks of social bots or exposed to disinformation. A notable example in~\cite{wagner2012social} builds on interactions (mentions, replies, retweets and friendship) between genuine and bot accounts, to obtain a ground truth of users susceptible to social bots. Then, similar to our approach, different learning algorithms have been adopted to train a classifier. Contrary to their approach, the current work is able to classify users close to social bots with lightweight features, all computed from data available in the user's profile.  Another brand new line of research is the detection of users susceptible to fake news. Work in~\cite{DBLP:conf/websci/ShenCGLYL19}  monitors the replies of Twitter users to a priori known fake news, in order to tag the same users as vulnerable to disinformation or not. Then, a supervised classification task is launched, to train a model able to classify gullible users, according to content-, user-, and network-based features.

\section{Conclusions}
\label{sec:conclusions}
\vspace{-0.5em}
Inspired by recent literature that shows how disinformation is not only promoted by social bots but also emphasized by genuine peers, in this work we proposed a supervised classification engine to discriminate {\it credulous} users, i.e., human-operated accounts with a high percentage of bots as friends. The classifier achieves very good performances and avoids a heavy feature engineering and extraction phase. Further research efforts will be devoted to investigate the behaviour of credulous users, as well as the posted content, to know more about their peculiarities and the quality of information they contribute to diffuse.

\bibliographystyle{splncs04}
\bibliography{biblio}

\end{document}